# G-quadruplexes and mRNA localization

Valentina Agoni
valentina.agoni@unipv.it

**G-quadruplexes represent a novelty for molecular biology. Their role inside the cell remains mysterious.**
**We investigate a possible correlation with mRNA localization.**
**In particular, we hypothesize that G-quadruplexes influence fluid dynamics.**

Localization of mRNA molecules within the cytoplasm recently emerged as a prevalent mechanism for cell polarization, thus underlying developmental processes.
[1] reports 5 adventages for a cell of transporting mRNAs rather than proteins:
(i) transport costs are reduced, as several protein molecules can be translated from a single RNA
molecule
(ii) transporting mRNAs can prevent proteins from acting ectopically
(iii) localized translation can facilitate incorporation of proteins into macromolecular complexes [2]
(iv) nascent proteins may have properties distinct from pre-existing copies, by virtue of post-translational modifications or through chaperone-aided folding pathways [3]
(v) a major advantage of mRNA targeting is that it allows fine-tuning of gene expression in both space and time [4] [5].
Until now three distinct mechanisms have been identified underlying mRNA localization: localized protection from degradation, diffusion-coupled local entrapment and directed transport along a polarized cytoskeleton.
But what can control mRNA localization?
At the same time the role of fluid dynamics in developmental biology has been acknowledged [6]. Human conception, indeed fertilization in general, takes place in a fluid.
Plant roots direct their growth in response to gravity, light, and mechanical stimuli. Local changes in hormone concentration have been shown to be part of the response mechanisms that result in those tropic growth.
First of all it is very interesting to note that gravistimulation and mechanical stimulation act on the same transcripts [7].
Furthermore [8] indicates that salt-induced altered gravitropism of root growth is mediated by ion disequilibrium.
We know that objects with the same mass do fall at the same speed in the vacuum and hit the ground at the same time, but air resistance hinders this fact. For example, if you drop a tennis ball and a feather at the same time on the Earth, obviously the ball falls faster.
In addition we have to consider parasitic drag. Parasitic drag is caused by moving a solid object through a fluid medium (in the case of hydrodynamics, more specifically, a liquid medium). Parasitic drag is made up of many components, the most prominent being form drag. Skin friction and interference drag are also major components of parasitic drag. The shape of the object is the most important factor in form drag or pressure drag: bodies with a larger apparent cross-section will have a higher drag than thinner bodies. The fluid we are considering is cytoplasm while the objects of our analysis are mRNAs.
Cytoplasmic streaming, also termed cyclosis, is a persistent circulation of the fluid contents of cells and we assume it to flow in every direction [6]. However what happens if we consider a very short instant of time the flow goes in one direction or an higher gravity system? Well, we talk about the tennis ball and the feather in the presence of gravity, now we also have to consider that the shape of the object is the most important factor in form drag or pressure drag: bodies with a larger apparent cross-section will have a higher drag than thinner bodies.
In [7] root apices from 7-day-old etiolated *Arabidopsis* seedlings were harvested and analyzed for relative changes in transcript
levels in response to gravistimulation and mechanical stimulation using the *Arabidopsis*

ATH1 GeneChip (Affymetrix) in a time course experiment. The analysis of whole genome microarrays on the differential regulation of transcript abundance reveals changes during the first hour after gravity [7]. Four-stranded G-quadruplex nucleic acid structures are G-rich DNA and RNA sequences that can fold into non-canonical secondary structures [9].

G-quadruplexes are coded by different sequences (Figure 1) but we identify a common motif in the AGGG sequence.

| Sequence |
| --- |
| 5′-GG(TTAGGG)$_4$TTAG-3′ |
| 5′-GG(TTAGGG)$_4$TTAG-3′ /3′-C(AATCCC)$_4$AAT-5′ |
| 5′-GGCATAGTGCGTGGGCG-3′ |
| 5′-TGAGGGTGGGTAGGGTGGGTAA-3′ |
| 5′-AGGGAGGGCGCTGGGAGGAGGG-3′ |
| 5′-CGGGCGGGCGCGAGGGAGGGG-3′ |
| 5′-GGCGAGGAGGGGCGTGGCCGGC-3′ |
| 5′-GGTTGGTGTGGTTGG-3′ |
| 5′-CAGUACAGAUCUGUACUG-3′ |
| 5′-(TTAGGG)$_3$-3′ |

**Figure 1.** Oligonucleotides used in enzyme-linked immunosorbent assays. Modified from [9].

We considered the sequences of some mRNAs shown to coherently increase or decrease in response to gravity and movement [7] analyzing the frequency of AGGG sequences.

It emerges that increasing-mRNAs present a significantly lower AGGG content respect to decreasing-mRNAs.

This can be reasonable if we think that increasing gravity the number of G-quadruplexes in transcripts should be lower (on average). This is the reason why we performed a one tailed-test.

Nevertheless the two populations of transcripts are heterogeneous because also in these conditions there should be mRNAs at different levels in the cellular cytoplasm.

In consideration of all these punctualizations and due to the limited number of samples we performed a Mann-Whitney test for not-Gaussian populations. The results are shown in Table 1.

In conclusion mRNAs could arrange in the cell according to their weight and shape also in terms of presence and number of G-quadruplexes.

But is this the major phenomenon regulating mRNA localization?

| NCBI Reference Sequence | AGGG percentage | NCBI Reference Sequence | AGGG percentage |
| --- | --- | --- | --- |
| Sequences increasing with gravity | | Sequences decreasing with gravity | |
| NM_179725.1 | 0.041701418 | NM_117622.2 | 0.160578081 |
| DQ105587.1 | 0.07199424 | AY150484.1 | 0.244498778 |
| AF332444.1 | 0.258397933 | AY093066.1 | 0.155884645 |
| NM_117342.2 | 0.066934404 | BT004357.1 | 0.384615385 |
| BT002780.1 | 0.092936803 | AY096456.1 | 0.666666667 |
| NM_104261.2 | 0.145772595 | BT003121.1 | 0.225988701 |
| BT010322.1 | 0.174216028 | AY051075.1 | 0.199071002 |
| AY045865.1 | 0.186046512 | BT003667.1 | 1.089324619 |
| BT012546.1 | 0.228310502 | BT025023.1 | 0.366300366 |
| AK317670.1 | 0.291828794 | AY142594.1 | 0.185070944 |
| NM_116393.1 | 0.260416667 | AY081574.1 | 0.257069409 |
| BT020422.1 | 0.229885057 | AF361097.1 | 0.336021505 |
| average | 0.170703413 | | 0.355924175 |
| stdev | 0.085971011 | | 0.270824472 |
| Z-ratio | | | -2.0496 |
| U-value | | | 36 |

**Table 1.** The AGGG/sequence length percentages relative to genes responsive to gravity and mechanical stimulation [7] are reported.